\documentclass{ws-p9-75x6-50}

\begin{document}

\title{Surrogate data for non--stationary signals}

\author{Andreas Schmitz and Thomas Schreiber}

\address{Physics Department, University of Wuppertal,\\
      D--42097 Wuppertal, Germany}

\maketitle

\abstracts{ 
  Standard tests for nonlinearity reject the null hypothesis of a
  Gaussian linear process whenever the data is non--stationary. Thus,
  they are not appropriate to distinguish non--{\em linearity} from
  non--{\em stationarity}. We address the problem of generating proper
  surrogate data corresponding to the null hypothesis of an ARMA
  process with slowly varying coefficients.}

\section{Introduction}
Most methods used in the field of linear and nonlinear time series
analysis assume stationarity of the considered data. Non--stationarity
is very likely to lead to wrong results.  This is especially true for
tests for nonlinearity.  A common approach is to split the time series
into segments which can be considered nearly stationary and perform
individual tests. But for short time series or not too slowly varying
non--stationarities these segments have to be made too short to
meaningfully calculate a test statistic on them.

\section{Surrogate Data}  
To generate surrogate data that has the same time variation of
autocorrelations than the original data one can in principle follow
the general approach of constrained randomization.\cite{constraint}
The corresponding cost function would be the discrepancy between the
autocorrelations of the original and the surrogate data in sliding
windows. Autocorrelations should be calculated up to sufficiently high
lags. The most striking disadvantage of this method is the extreme
computational burden. Therefore, we use an alternative method:\\

\begin{tabular}{ll}
1. & split up the time series into segments\\
2. & \label{zwei} generate a surrogate for each segment\\
3. & join the segments to one new surrogate\\
\end{tabular}\\

For step \ref{zwei} we use an iterative
algorithm.\cite{surrowe,tisean} The size of the segments is typically
too small to perform individual tests for nonlinearity on them,
because most test statistics need sufficiently large data sets. A
disadvantage of the method is the loss of correlations {\em between}
the segments which can lead to a bias.

\section{Cyclostationary Processes}
The first class of processes we consider are cyclostationary, having
periodically varying parameters. These processes have been found to
yield rejections of the null hypothesis\cite{timmer} when using a test
statistic derived from the correlation sum. Here we use the maximum
likelihood estimator of the correlation
dimension\cite{takens,theiler88,tisean}
\begin{equation}
\label{eq:takens}
D_2^{\rm ML}={C(\epsilon_0) \over 
  \int\limits_0^{\epsilon_0}{C(\epsilon) / \epsilon} \; d\epsilon}
\, .
\end{equation}
Following Timmer\cite{timmer}, the considered time series is generated
by an AR(2)--process
\begin{equation}
  \label{eq:AR2}
  x_{n} = a_1 x_{n-1} + a_2 x_{n-2} + \eta_{n}, \quad
  n = 1,\ldots,2000
\end{equation}
with coefficients chosen to be
\begin{eqnarray}
  \label{eq:a1a2}
  a_1 & = & 2\cos(2\pi/T)\exp(-1/\tau)\\
  a_2 & = & -\exp(-2/\tau) \nonumber
\end{eqnarray}
where $T$ is the period and $\tau$ the relaxation time of a damped
oscillator. We will use $T=7$ and $\tau=50$. Non--stationarity is
introduced by a modulation of the period $T$
\begin{equation}
  \label{eq:modula}
  T(n)=T_{\rm mean}+A\sin(2\pi/T_{\rm mod} n)
\, .
\end{equation}
By varying $A$ and $T_{\rm mod}$ the degree and time scale of the
non--stationarity can be changed which is reflected in non--constant
running variances and autocorrelations.

\subsection{Testing the test size}  
The fraction of tests that falsely reject the null hypothesis, when it
is in fact true, is called the {\em size} of the test. Since a single
test\cite{timmer} says not much about the size, we performed $1000$
one--sided tests with $9$ surrogates for each process. A correct test
with proper surrogates should yield the nominal size $\alpha=0.1$.

\begin{figure}[t]
\begin{center}
\epsfbox{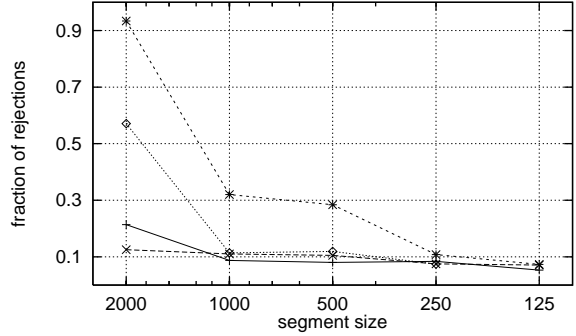}
\end{center}
\caption{Size of nonlinearity tests with varying segment sizes. The
  modulation period of the non--stationarity is fixed to $T_{\rm
    mod}=2000$. From upper to lower curve the values $A=$2.0, 1.5, 1.0
  and 0.0 are used.}
\label{fig:s2000}
\end{figure}

For the first tests we used $T_{\rm mod}=2000$ for the modulation
period. As shown in Fig.~\ref{fig:s2000} the new method does indeed
yield the correct size of the test for a segment size of about $T_{\rm
  mod}/2=1000$. Only for relatively strong non--stationarity ($A=2.0$)
smaller segments seem to be necessary.
\begin{figure}[t]
  \begin{center}
\epsfbox{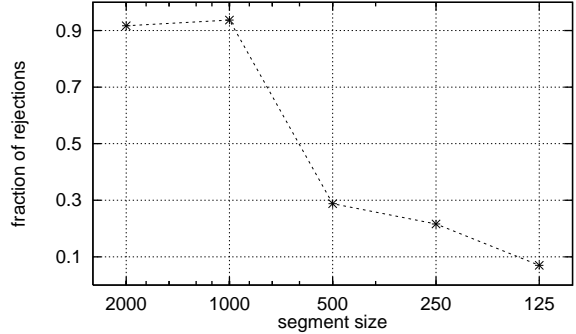}
  \end{center}
\caption{Size of nonlinearity tests with varying segment sizes. Now
  $T_{\rm mod}=1000$ and $A=2.0$ are used.}
\label{fig:s1000}
\end{figure}
If we reduce the time scale of the non--stationarity by setting
$T_{\rm mod}=1000$ we get sizes shown in Fig.~\ref{fig:s1000}. As
expected, only segment sizes which are smaller than $T_{\rm mod}$ can
decrease the number of false rejections.

\section{AR(1)--Processes}
We found cases in which non--stationarity does not yield a higher
rate of rejections of the null hypothesis but just the opposite.
Consider an AR(1)-process 
\begin{equation}
  \label{eq:AR1}
  x_{n}=a_1 x_{n-1} + \eta_{n}
\end{equation}
with a nonlinear measurement function $M(\cdot)$
\begin{equation}
  \label{eq:Mx}
  M(x_{n})=x_n\sqrt{|x_n|}
\, .
\end{equation}
Non--stationarity is introduced by varying $a_1$ from $0.5$ to $0.9$
along the time series. We now choose a {\em prediction error} as test
statistic.\cite{tisean} It is based on locally constant predictions in
an $m$--dimensional embedding space.

\begin{figure}[t]
\epsfbox{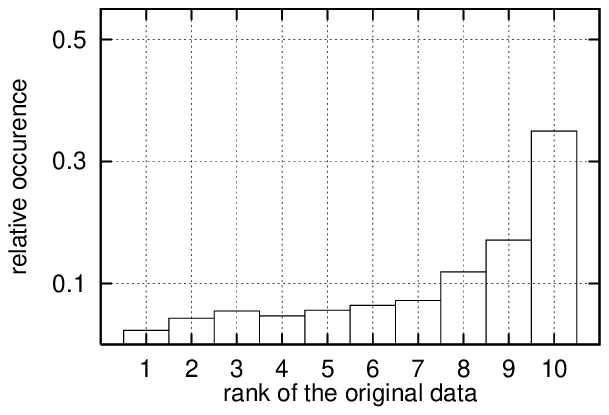}
\epsfbox{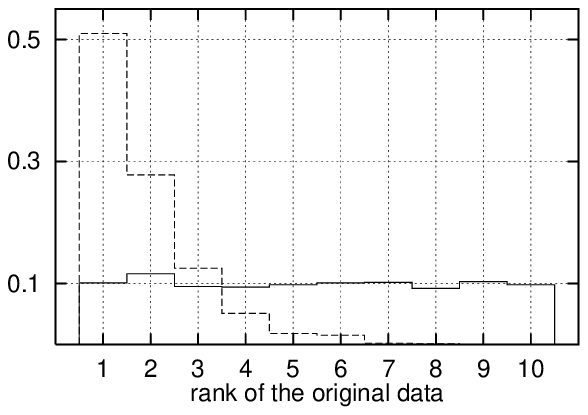}
  \caption{Distributions of the rank of the original data within the
    surrogates for stationary surrogates (left) and surrogates with
    segment sizes 1000 (solid) and 250 (dashed).}
  \label{fig:histall}
\end{figure}

Again we performed 1000 one--sided tests with 9 surrogates to estimate
the size.  Additionally we calculated the rank of the original data
within the surrogates going from 1 if the prediction error of the
original data set was lower than for all surrogates up to 10 if the
original data yields the largest prediction error. For usual
stationary surrogates we get the distribution presented on the left
side of Fig.~\ref{fig:histall}. In 35\% of the tests the original data
has a {\em larger} prediction error than all the surrogates and the
null is rejected in only 2.3\% of the tests. This is the opposite of
what we expect for non--{\em linear} signals, which should be better
predictable. Here, predictions are easier for the stationary
surrogates with constant autocorrelations than for the original
non--stationary data. Prediction error is not a useful discriminating
statistic in this case.

On the right side of Fig.~\ref{fig:histall} the results for tests with
the new surrogates and different segment sizes are shown. For a
segment size of $1000$ we get the correct size and therefore a uniform
distribution. With decreasing segment sizes the prediction error of
the surrogates gets larger than for the original data. The predictions
seem to be very sensitive to the autocorrelations that get lost
between the segments.

\section{IV. Summary}
Dealing with non--stationary signals is still a difficult business. We
have given numerical evidence that in a test for nonlinearity of a
data set with possible non--stationarity in form of slow variation it
may be feasible to generate surrogates on separate segments.  These
segments can be made smaller than would be necessary for individual
nonlinearity tests. But the segment size and the test statistic should
still be chosen with great care, since no general theory guarantees
the correctness of the test. As a final remark we want to stress that
the reasoning in this work is not applicable to sudden changes like
jumps or spikes.

\end{document}